\documentclass[12pt]{article}
\usepackage[utf8]{inputenc}
\usepackage{amsmath,setspace,geometry}
\usepackage{amsthm}
\usepackage{amsfonts}
\usepackage[shortlabels]{enumitem}
\usepackage{rotating}
\usepackage{pdflscape}
\usepackage{graphicx}
\usepackage{bbm}
\usepackage[dvipsnames]{xcolor}
\usepackage{hyperref}
\hypersetup{colorlinks=true, linkcolor= BrickRed, citecolor = BrickRed, filecolor = BrickRed, urlcolor = BrickRed, hypertexnames = true}
\usepackage[]{natbib} 
\bibpunct[:]{(}{)}{,}{a}{}{,}
\usepackage[english]{babel}
\usepackage{float}
\usepackage{caption}
\usepackage{subcaption}
\usepackage{booktabs}
\usepackage{pdfpages}
\usepackage{threeparttable}
\usepackage{lscape}
\usepackage{bm}
\usepackage{multirow,bigdelim}
\bibpunct[:]{(}{)}{,}{a}{}{,}
\newtheorem{proposition}{Proposition}

\usepackage{setspace}
\begin{document}
\title{A Note on Identification of Match Fixed Effects as Interpretable Unobserved Match Affinity}
\author{Suguru Otani\thanks{suguru.otani@e.u-tokyo.ac.jp, Market Design Center, University of Tokyo}, Tohya Sugano\thanks{sugano-tohya1011@g.ecc.u-tokyo.ac.jp, University of Tokyo\\Declarations of interest: none} }
\maketitle

\begin{abstract}
\noindent
We highlight that match fixed effects, represented by the coefficients of interaction terms involving dummy variables for two elements, lack identification without specific restrictions on parameters. Consequently, the coefficients typically reported as relative match fixed effects by statistical software are not interpretable. To address this, we establish normalization conditions that enable identification of match fixed effect parameters as interpretable indicators of unobserved match affinity, facilitating comparisons among observed matches.
Using data from middle school students in the 2007 Trends in International Mathematics and Science Study (TIMSS), we highlight the distribution of comparable match fixed effects within a specific school.
\textbf{Keywords}: Match fixed effect, Affinity, Identification \\
\textbf{JEL code}: C21, J24, J31, I26
\end{abstract}

\section{Introduction}
Evaluating the quality of matching and affinity between entities is common in empirical research. Affinity is divided into observed and unobserved components. Observed affinity is typically measured by the coefficient of interaction terms involving observable characteristics, while unobserved affinity is captured through match fixed effects using dummy variables. This approach is widely used in labor and education economics to analyze and interpret match quality across pairs such as teacher-student and worker-company relationships. Our study aims to clarify the identification process of match fixed effects.

As an illustrative example, \cite{inoue2023teachers} investigate the impact of a teacher's major on students' achievement using the following econometric model:
\begin{align}
    Y_{ifj} = \beta Major_{fj} + \delta_f + \eta_{ij} + \varepsilon_{ifj}, \label{eq:inoue_tanaka_specification}
\end{align}
where \( Y_{ifj} \) represents the science test score of student \( i \) in subfield \( f \) within class \( j \). The parameter \( \beta \) denotes the coefficient of interest, and \( Major_{fj} \) is an indicator variable that indicates whether the teacher's major field in natural science matches the subfield of the student's test score. \( \delta_{f} \) denotes the fixed effect specific to subfield \( f \), while \( \eta_{ij} \) represents the student-teacher fixed effects, accounting for any subfield-invariant determinants of science test scores between student \( i \) and teacher \( j \), thereby capturing their unobserved affinity. The authors find a significant increase in R-squared upon introducing student-teacher fixed effects, underscoring the importance of match fixed effects in their analysis.

Similar methodologies are utilized in various studies, including pitcher-catcher fixed effects on strikeout likelihood \citep{biolsi2022task}, worker-company fixed effects on income \citep{mittag2019simple} and turnover rates \citep{ferreira2011measuring}, student-school fixed effects on student performance \citep{ovidi2022parents}, teacher-school fixed effects on student test scores \citep{jackson2013match}, and student-university fixed effects on post-graduation income \citep{dillon2020consequences}. Additionally, some studies account for two-way fixed effects by separately considering the fixed effects on each side.

Despite their common use, the interpretation of match fixed effects as affinity indicators for all matches remains unclear. Our paper addresses this by distinguishing between relative and absolute match fixed effects within a standard model. We argue that absolute match fixed effects are not identifiable without specific restrictions, whereas relative match fixed effects are identifiable. We propose location normalization conditions to enable the identification and interpretation of absolute match fixed effects, facilitating comparisons of unobserved match quality.

Applying this approach to the data from \cite{inoue2023teachers}, we demonstrate the distribution of unobserved match quality between students and teachers in a specific school. Our analysis reveals that one teacher excels with high-achieving students but underperforms with lower-achieving ones, while another teacher shows the opposite pattern in terms of unobserved match affinity.

\section{Model}
We consider the following typical setting.
Suppose that we can observe $I$ students indexed by $i$ and $J$ teachers indexed by $j$ at time $t=1,\cdots,T$.
Student $i$ makes some outcome with teacher $j$ at time $t$.
For avoiding later notational complexity, we do not include time fixed effects like a panel regression, but the inclusion does not affect our findings.
We consider the following regression model:
\begin{align}
Y_{ijt}&=\alpha_{i}+\beta_{j}+\mu_{ij}+ X_{ijt}'\gamma+\varepsilon_{ijt}, \label{eq:orignal_regression}\\
&=\sum_{i'=1}^{I}\alpha_{i'}1(i=i')+\sum_{j'=1}^{J}\beta_{j'}1(j=j')+\sum_{i'=1}^{I}\sum_{j'=1}^{J}\mu_{i'j'}1(i=i',j=j')+ X_{ijt}'\gamma+\varepsilon_{ijt}\label{eq:matrix_regression}
\end{align}
where $Y_{ijt}$ is the test score of student $i$ with teacher $j$ at time $t$, $X_{ij}$ is $d$-dimensional covariates consisting of observed characteristics of student $i$ and teacher $j$ and its interaction at time $t$, $\alpha_{i}$ is student $i$'s fixed effect, $\beta_{j}$ is teacher $j$'s fixed effect, and $\mu_{ij}$ is $(i,j)$-match fixed effect which is of our interest, $\gamma$ is a $d$-dimensional vector of parameters, $1(\cdot)$ is an indicator function, and $\varepsilon_{ijt}$ is an error term assumed to be drawn i.i.d from standard normal distribution. 
Note that we explicitly decompose affinity between student $i$ and teacher $j$ into two parts, that is, $X_{ijt}'\gamma$ and $\mu_{ij}$ as the observed and unobserved affinities.
For later discussion, we call $\mu_{ij}$ the \textit{absolute} match effect of student $i$ and teacher $j$. Similarly, we call $\alpha_{i}$ and $\beta_{j}$ the absolute fixed effects.\footnote{Using Equation \eqref{eq:matrix_regression}, matrix representation is described as $Y=X\gamma+1_{I}\alpha+1_{J}\beta+1_{IJ}\mu+\varepsilon=X\gamma+\tilde{X}\delta + \varepsilon$
where $Y$ is $IJT\times 1$, $X$ is $IJT\times d$, $\gamma$ is $d\times 1$, $1_{I}$ is $IJT\times I$, $\alpha$ is $I\times 1$, $1_{J}$ is $IJT\times J$, $\beta$ is $J\times 1$, $1_{IJ}$ is $IJT\times IJ$, $\mu$ is $IJ\times 1$, $\varepsilon$ is $IJT\times 1$, and denote $\tilde{X}=[1_{I} 1_{J} 1_{IJ}]$ and $\delta=[\alpha^T \beta^T \mu^T]^T$. Then, define $I$ as $IJT\times 1$ one vector and $M=I-\tilde{X}(\tilde{X}^T\tilde{X})^{-1}\tilde{X}^{T}$ as the annihilator matrix for $\tilde{X}$. The OLS estimator of $\delta$ is obtained as $\hat{\delta}=(\tilde{X}^T M\tilde{X})^{-1}(\tilde{X}^T MY)$, so the standard full rank condition for $\delta$ is $Rank(\tilde{X}^T M\tilde{X})=(I+J+IJ)$. See \cite{hansen2022econometrics} Chapter 3.16 for reference. However, the condition does not hold due to multicollinearity.}

In the context of the standard fixed effect model, when there are \( I \) groups, typically \( I - 1 \) fixed effects are incorporated, alongside a constant term, to avoid multicollinearity that would arise from including the \( I \)-th group. The match fixed effect case is more complex. Similarly, for a student \( i \neq 1 \) and teacher \( j \neq 1 \), Equation \eqref{eq:orignal_regression} can be rewritten as
\begin{align}
    Y_{ijt}&=\underbrace{\alpha_{1}+\beta_{1}+\mu_{11}}_{\text{constant}}\nonumber\\
    &+\underbrace{(\beta_{j}-\beta_{1})+(\mu_{1j}-\mu_{11})}_{\text{teacher }j\text{'s relative fixed effect}}\nonumber\\
    &+\underbrace{(\alpha_{i}-\alpha_{1})+(\mu_{i1}-\mu_{11})}_{\text{student }i\text{'s relative fixed effect}}\nonumber\\
    &+\underbrace{(\mu_{ij}+\mu_{11}-\mu_{1j}-\mu_{i1})}_{(i,j)\text{'s relative match effect}}+ X_{ijt}'\gamma+\varepsilon_{ijt},
\end{align}
where the first line is a constant parameter normalized to student 1 and teacher 1 which are arbitrarily chosen, the second line is called teacher $j$'s \textit{relative} fixed effect which is the fixed effect relative to teacher 1's fixed effects, the third line is called student $i$'s \textit{relative} fixed effect which is the fixed effect relative to student 1's fixed effects, the third line is called the \textit{relative} match effect of student $i$ and teacher $j$ relative to student 1 and teacher 1.
Avoiding multicollinearity, we can identify and estimate these relative fixed effects and $\gamma$ instead of absolute fixed effects. 
Statistical software automatically reports the estimates of the relative fixed effects.

Relative fixed effects indicate how a match compares to a specific reference match, rather than categorizing it as "good" or "bad" compared to all other matches. For controlling match fixed effects or obtaining overall affinity without distinguishing observed from unobserved components, relative fixed effects are adequate. However, for measuring and understanding unobserved affinity, such as personality match quality, relative fixed effects are insufficient. In these cases, absolute fixed effects are crucial for meaningful comparison and interpretation of unobserved affinity.

Our central question posed is: ``Can we derive the absolute fixed effects from the estimated relative fixed effects without imposing any restrictions?" In mathematical terms, ``Can we solve the system of equations involving relative fixed effects for absolute fixed effects without restrictions?"
Our conclusion is in the negative: No, we cannot achieve this without imposing constraints.
Subsequently, we propose location normalization restrictions that are necessary for identifying the absolute fixed effects \( \mu_{ij} \), summarized in Proposition \ref{prop:main_results}.
\begin{proposition}\label{prop:main_results}
For regression model \eqref{eq:orignal_regression}, the following results hold.
\begin{enumerate}
    \item Without any restriction, the absolute match effect $\mu_{ij}$ is not identified. 
    \item With $\sum_i \mu_{i j}=\sum_j \mu_{i j}=0$, the absolute match effect $\mu_{ij}$ for all $i$ and $j$ is identified by relative fixed effects. 
    \item With restriction $\sum_i \mu_{i j}=\sum_j \mu_{i j}=\sum_i \alpha_i=\sum_j \beta_j=0$ for all $i$ and $j$, the absolute fixed effects $\alpha_i$, $\beta_j$, and absolute match effect $\mu_{ij}$ for all $i$ and $j$ are identified by relative fixed effects.
\end{enumerate}

\end{proposition}
See the proof and illustrative example in Appendix \ref{sec:proof}.
Intuitively, the conditions outlined are effective for location normalization. The restricted match effect indicates how much better the match is compared to the average match, rather than a specific match. Consequently, it yields positive values when the match is relatively superior and negative values when it is relatively inferior compared to the average match. This approach ensures interpretability across students and teachers, facilitating meaningful comparisons of unobserved affinity.

\section{Empirical exercise}

\begin{figure}[!ht]
\begin{center}
\includegraphics[height = 0.55\textheight]{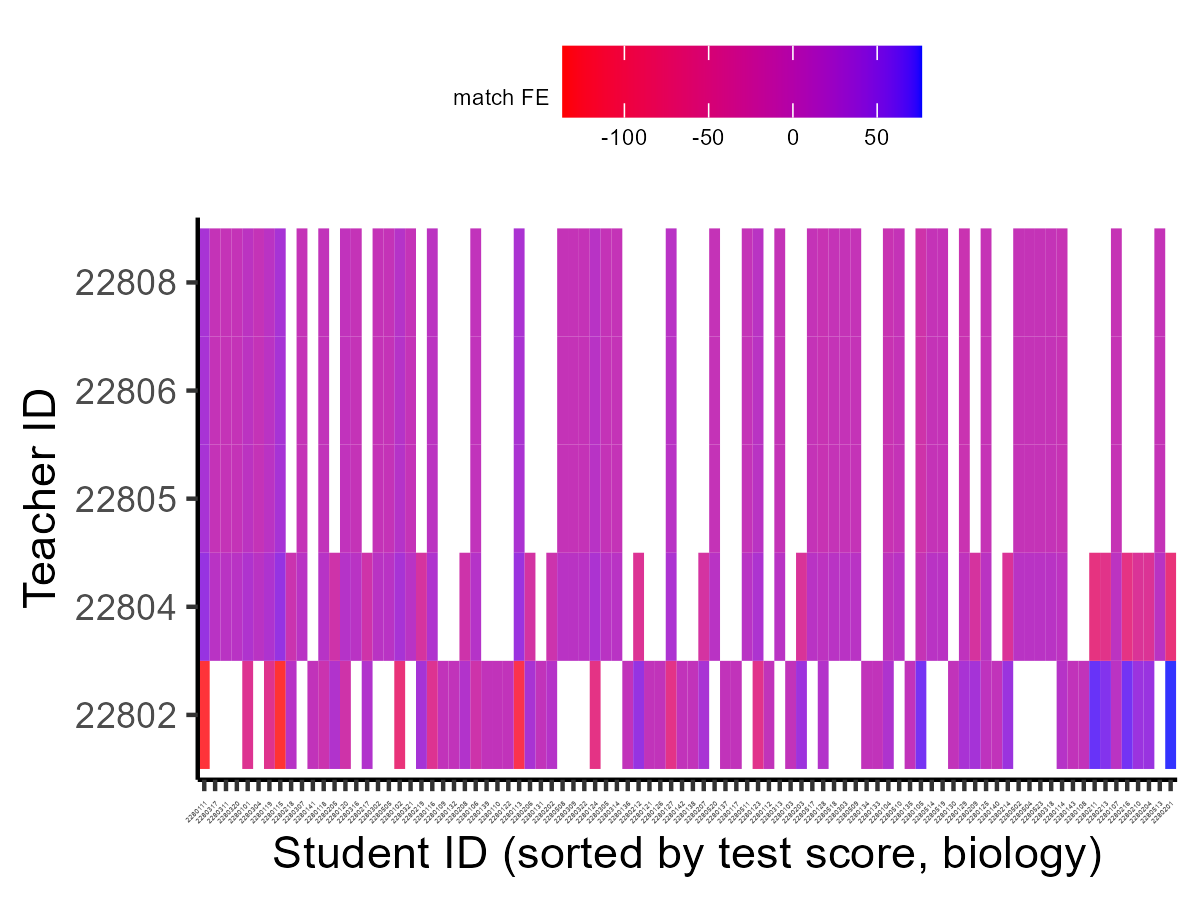}
\end{center}
\caption{Heatmap of absolute match fixed effects}\footnotesize
\textit{Notes}: 
The blank cell shows no match in the data. The estimated coefficient of the major match in Equation \eqref{eq:inoue_tanaka_specification} is 497.56 (Standard error: 8.0).
\label{fg:heatmap_match_effect}
\end{figure}

To illustrate our approach, we use data from the 2007 Trends in International Mathematics and Science Study (TIMSS), as in \cite{inoue2023teachers}. This dataset includes middle school students' test scores in various science subfields (physics, chemistry, biology, and Earth science) and teacher-related variables. For details, refer to \cite{inoue2023teachers}. We focus on data from a specific school (ID 228) with five teachers and 90 students, using Equation \eqref{eq:inoue_tanaka_specification} to estimate absolute match fixed effects with location normalization. We exclude subfield variables to avoid multicollinearity with student-teacher interaction dummies, as some subjects are taught exclusively by certain teachers. The author's GitHub page provides Monte Carlo simulation code and results, as well as replication files for our empirical exercises.

Figure \ref{fg:heatmap_match_effect} presents a heatmap of absolute match fixed effects. Teacher ID 22802 shows better effects with high-performing students but poorer effects with low-performing students, while teacher ID 22804 displays the opposite pattern. This suggests that teacher 22802 excels with high achievers but underperforms with lower achievers, whereas teacher 22804 performs better with lower achievers. Furthermore, absolute match effects offer a more comparable measure than relative match effects. For instance, the absolute match effect for student ID 2280201 and teacher ID 22802 is 76.34, compared to 44.57 for student ID 2280115 and teacher ID 22804, indicating that the former match has 1.75 times higher unobserved affinity. Thus, absolute fixed effects are crucial for meaningful comparisons and interpretations of unobserved affinity.

\section{Conclusion}
We examine the estimation of affinities using match fixed effects, distinguishing between observed and unobserved components. We emphasize that, without proper restrictions, only relative fixed effects—fixed effects relative to normalized values—can be estimated, which do not provide interpretable measures of unobserved affinity. To enable meaningful interpretation of absolute match effects, we introduce theoretical constraints on parameters. Using 2007 TIMSS data on middle school students, we illustrate the distribution of absolute match fixed effects and underscore their significance.

\paragraph{Acknowledgments}
We thank Shunya Noda, Shosei Sakaguchi, and Ryuichi Tanaka for their valuable advice. This work was supported by JST ERATO Grant Number JPMJER2301, Japan.

\bibliographystyle{ecca}
\bibliography{match_effect}

\begin{thebibliography}{8}
\providecommand{\natexlab}[1]{#1}

\bibitem[{Biolsi \textit{et~al.}(2022)Biolsi, Goff and Wilson}]{biolsi2022task}
\textsc{Biolsi, C.}, \textsc{Goff, B.} and \textsc{Wilson, D.} (2022). Task-level match effects and worker productivity: evidence from pitchers and catchers. \textit{Applied Economics}, \textbf{54}~(25), 2888--2899.

\bibitem[{Dillon and Smith(2020)}]{dillon2020consequences}
\textsc{Dillon, E.~W.} and \textsc{Smith, J.~A.} (2020). The consequences of academic match between students and colleges. \textit{Journal of Human Resources}, \textbf{55}~(3), 767--808.

\bibitem[{Ferreira and Taylor(2011)}]{ferreira2011measuring}
\textsc{Ferreira, P.} and \textsc{Taylor, M.} (2011). Measuring match quality using subjective data. \textit{Economics Letters}, \textbf{113}~(3), 304--306.

\bibitem[{Hansen(2022)}]{hansen2022econometrics}
\textsc{Hansen, B.} (2022). \textit{Econometrics}. Princeton University Press.

\bibitem[{Inoue and Tanaka(2023)}]{inoue2023teachers}
\textsc{Inoue, A.} and \textsc{Tanaka, R.} (2023). Do teachers’ college majors affect students’ academic achievement in the sciences? a cross-subfields analysis with student-teacher fixed effects. \textit{Education Economics}, \textbf{31}~(5), 617--631.

\bibitem[{Jackson(2013)}]{jackson2013match}
\textsc{Jackson, C.~K.} (2013). Match quality, worker productivity, and worker mobility: Direct evidence from teachers. \textit{Review of Economics and Statistics}, \textbf{95}~(4), 1096--1116.

\bibitem[{Mittag(2019)}]{mittag2019simple}
\textsc{Mittag, N.} (2019). A simple method to estimate large fixed effects models applied to wage determinants. \textit{Labour Economics}, \textbf{61}, 101766.

\bibitem[{Ovidi(2022)}]{ovidi2022parents}
\textsc{Ovidi, M.} (2022). \textit{Parents Know Better: Sorting on Match Effects in Primary School}. Tech. rep., Universit{\`a} Cattolica del Sacro Cuore, Dipartimenti e Istituti di Scienze~….

\end{thebibliography}

\newpage
\appendix
\section{Proof (Online appendix)}\label{sec:proof}
\begin{proof}
    Before showing the proof, we provide the overview. The system of linear equations about parameters to be solved without any restrictions is
    \begin{equation*}
        \begin{pmatrix}
            A \\ B \\ C
        \end{pmatrix}
        \begin{pmatrix}
            \mathbf{\alpha} \\ \mathbf{\beta} \\ \mathbf{\mu}
        \end{pmatrix}=
        \begin{pmatrix}
            \mathbf{\alpha'} \\ \mathbf{\beta'} \\ \mathbf{\mu'}
        \end{pmatrix}
    \end{equation*}
    where $A, B$ and $C$ are the coefficient matrix for relative effects and $\alpha',\beta'$ and $\mu'$ are calculated relative fixed effects defined later.
    Let $T$ denote transpose.
    Then if $rank\begin{pmatrix}
        [A^T&B^T&C^T]^T
    \end{pmatrix}=rank\begin{pmatrix}
        [\alpha^T&\beta^T&\mu^T]^T
    \end{pmatrix}=I+J+IJ$, the absolute fixed effects and match effects are just identified. Thus, we will check the rank conditions to determine whether a system of linear equations is underdetermined (i.e., underidentified), meaning there are fewer equations than unknowns.\\
    Since there are $I$ students and $J$ teachers, there are $I+J$ unknown absolute fixed effect parameters and $IJ$ unknown absolute match effect parameters. The fixed effects of $(i,j)$ relative to $(i_0,j_0)$, $\alpha'_{i_0j_0,i},\beta'_{i_0j_0,j},\mu'_{i_0j_0,ij}$ is calculated by 
    \begin{align}
        \alpha'_{i_0j_0,i} &= \alpha_i-\alpha_{i_0}+\mu_{ij_0}-\mu_{i_0j_0}\label{eq:relative_alpha} \\
        \beta'_{i_0j_0,j} &= \beta_j-\beta_{j_0}+\mu_{i_0j}-\mu_{i_0j_0} \label{eq:relative_beta} \\
        \mu'_{i_0j_0,ij} &= \mu_{ij}+\mu_{i_0j_0}-\mu_{ij_0}-\mu_{i_0j}. \label{eq:relative_mu} 
    \end{align}
    First, we prove that absolute match effects are not identifiable without any restrictions.
    Let coefficient matrices $A$, $B$ and $C$ represent coefficient of $\alpha_i$, $\beta_j$ and $\mu_{ij}$ in Equations \eqref{eq:relative_alpha}, \eqref{eq:relative_beta}, and \eqref{eq:relative_mu} respectively as follows:
    \arraycolsep=4pt
    \begin{equation*}
        \begin{array}{crcccccccccccccl}
            \multirow{4}{*}{A\ =}&\ldelim({4}{14pt}[] & a^{11,2}_{1} & \cdots & a^{11,2}_{I} & 0 & \cdots & 0 & c^{11,2}_{11} & \cdots & c^1_{1J} & \cdots & c^{11,2}_{I1} & \cdots & c^{11,2}_{IJ} & \rdelim){4}{14pt}[] \\
            && a^{11,3}_{1} & \cdots & a^{11,3}_{I} & 0 & \cdots & 0 &c^{11,3}_{11} & \cdots & c^{11,3}_{1J} & \cdots & c^{11,3}_{I1} & \cdots & c^{11,3}_{IJ} & \\
            && \vdots & \vdots & \vdots & \vdots & \vdots & \vdots & \vdots & \vdots &\vdots & \vdots & \vdots & \vdots & \vdots &\\
            && a^{IJ,I-1}_{1} & \cdots & a^{IJ,I-1}_{I} & 0 & \cdots & 0 &c^{IJ,I-1}_{11} & \cdots & c^{IJ,I-1}_{1J} & \cdots & c^{IJ,I-1}_{I1} & \cdots & c^{IJ,I-1}_{IJ} &\\
            \multirow{4}{*}{B\ =}&\ldelim({4}{14pt}[] & 0 & \cdots & 0 & b^{11,2}_{1} & \cdots & b^{11,2}_{J} & c^{11,2}_{11} & \cdots & c^{11,2}_{1J}  & \cdots & c^{11,2}_{I1} & \cdots & c^{11,2}_{IJ} & \rdelim){4}{14pt}[] \\
            && 0 & \cdots & 0 & b^{11,3}_{1} & \cdots & b^{11,3}_{J} &c^{11,3}_{11} & \cdots & c^{11,3}_{1J}  & \cdots & c^{11,3}_{I1} & \cdots & c^{11,3}_{IJ} & \\
            && \vdots & \vdots & \vdots & \vdots & \vdots & \vdots & \vdots & \vdots & \vdots &\vdots & \vdots & \vdots & \vdots &\\
            && 0 & \cdots & 0 & b^{IJ,J-1}_{1} & \cdots & b^{IJ,J-1}_{J} &c^{IJ,J-1}_{11} & \cdots & c^{IJ,J-1}_{1J}  & \cdots & c^{IJ,J-1}_{I1} & \cdots & c^{IJ,J-1}_{IJ} &\\
            \multirow{4}{*}{C\ =}&\ldelim({4}{14pt}[] & 0 & \cdots & 0 & 0 & \cdots & 0 & c^{11,2}_{11} & \cdots & c^{11,2}_{1J}  & \cdots & c^{11,2}_{I1} & \cdots & c^{11,2}_{IJ} & \rdelim){4}{14pt}[] \\
            && 0 & \cdots & 0 & 0 & \cdots & 0 &c^{11,3}_{11} & \cdots & c^{11,3}_{1J}  & \cdots & c^{11,3}_{I1} & \cdots & c^{11,3}_{IJ} & \\
            && \vdots & \vdots & \vdots & \vdots & \vdots & \vdots & \vdots & \vdots & \vdots &\vdots & \vdots & \vdots & \vdots &\\
            && 0 & \cdots & 0 & 0 & \cdots & 0 &c^{IJ,J-1}_{11} & \cdots & c^{IJ,J-1}_{1J}  & \cdots & c^{IJ,J-1}_{I1} & \cdots & c^{IJ,J-1}_{IJ} &\\
             && \multicolumn{13}{l}{\underbrace{\hspace{7.5em}}_{I}\hspace{1.5em}\underbrace{\hspace{7.5em}}_{J}\hspace{1.5em}\underbrace{\hspace{18em}}_{IJ}} & \\
        \end{array}
    \end{equation*}
    \arraycolsep=5pt
    where $a^{i_0j_0,i}_{k}$ and $b^{j_0j_0,i}_{k}$ represent the coefficients of $\alpha_k$ and $\beta_k$, and $c^{i_0j_0,i}_{kl}$ represents the coefficient of $\mu_{kl}$ in the equation $\mu'_{i_0j_0,ij}=\alpha_i-\alpha_{i_0}+\mu_{ij_0}-\mu_{i_0j_0}$. 
    Every element in each matrix is either $0, 1$ or $-1$.
    
    First, consider the matrix $C$. 
    Assume $c^{i_0j_0,ij}_{kl}=1\ (k\neq i_0,\ l\neq j_0)$, then $c^{i_0j_0,ij}_{i_0j_0}=1$, $c^{i_0j_0,ij}_{i_0l}=c^{i_0j_0,ij}_{kj_0}=-1$ and $c^{i_0j_0,ij}_{k'l'}=0$ for $k'\notin \{k,i_0\},\ l'\notin \{l,j_0\}$. 
    Then, $c^{ij,i_0j_0}=c^{i_0j_0,ij}$, $c^{ij_0,i_0j}=-c^{i_0j_0,ij}$, and $c^{i_0'j_0',ij}=c^{i_0j_0,ij}+c^{i_0j_0,i_0'j_0'}-c^{i_0j_0,i_0'j}-c^{i_0j_0,ij_0'}$ hold. 
    Here fix $i_0=j_0=1$. 
    Then $c^{i_0'j_0',i'j'}$ can be obtained from $c^{11,ij}$. 
    Thus, $rank(C)=(I-1)(J-1)$.
    Similarly, because $a^{i_0j_0,i}=a^{ij_0,i_0}$ and $a^{i_0'j_0',i}=a^{11,i}-a^{11,i_0'}+c^{ij_0',i_0'1}$, $rank
    \begin{pmatrix}
        [A^T& C^T]^T
    \end{pmatrix} = 
    (I-1)+(I-1)(J-1)=(I-1)J$.
    Also, because $b^{i_0j_0,j}=b^{i_0j,j_0}$ and $b^{i_0'j_0',j}=b^{11,j}-b^{11,j_0'}+c^{i_0'j,1j_0'}$, $rank\begin{pmatrix}
        [A^T & B^T & C^T]^T
    \end{pmatrix}= (J-1)+(I-1)J=IJ-1$.
    
    This concludes that $rank\begin{pmatrix}         [A^T & B^T & C^T]^T     \end{pmatrix}=IJ-1<IJ$ and absolute match effects are not identifiable. 
    Note that without match fixed effect, then all elements in $C$ is $0$, $rank\begin{pmatrix}
        [A^T & B^T]^T
    \end{pmatrix}=I+J-2$, and thus $\alpha$ and $\beta$ is not identifiable without additional restrictions. 
    
    Next, we prove that absolute match effects are identifiable with restrictions $\sum_i\mu_{ij}=\sum_j\mu_{ij}=1$.
    Each restriction is independent from every equation above and the number of independent restrictions is $I+J-1$. 
    Therefore, $rank\begin{pmatrix}
        [C^T & R_1^T]^T
    \end{pmatrix}=IJ$ where $R_1$ is the coefficient matrix for the restrictions $\sum_i\mu_{ij}=\sum_j\mu_{ij}=1$ as
    \begin{equation*}
        R_1 = 
        \begin{array}{rccccccccccccccccll}
            & \multicolumn{16}{l}{\hspace{10em}\overbrace{\hspace{16em}}^{IJ}}&&\\
            \ldelim({7}{14pt}[] & 0 & \cdots & 0 & 0 & \cdots & 0 & 1 & \cdots & 1 & 0 & \cdots & 0 & \cdots & 0 & \cdots & 0 & \rdelim){7}{14pt}[]&\rdelim\}{4}{14pt}[I]\\
            & 0 & \cdots & 0 & 0 & \cdots & 0 & 0 & \cdots & 0 & 1 & \cdots & 1 & \cdots & 0 & \cdots & 0 & & \\
            & \vdots & \vdots & \vdots & \vdots & \vdots & \vdots & \vdots & \vdots & \vdots & \vdots & \vdots & \vdots & \vdots & \vdots & \vdots & \vdots & &\\
             & 0 & \cdots & 0 & 0 & \cdots & 0 & 0 & \cdots & 0 & 0 & \cdots & 0 & \cdots & 1 & \cdots & 1 & &\\
              & 0 & \cdots & 0 & 0 & \cdots & 0 & 1 & \cdots & 0 & 1 & \cdots & 0 & \cdots & 1 & \cdots & 0 & & \rdelim\}{3}{14pt}[J]\\
              & \vdots & \vdots & \vdots & \vdots & \vdots & \vdots & \vdots & \vdots & \vdots & \vdots & \vdots & \vdots & \vdots & \vdots & \vdots & \vdots & &\\
              & 0 & \cdots & 0 & 0 & \cdots & 0 & 0 & \cdots & 1 & 0 & \cdots & 1 & \cdots & 0 & \cdots & 1 & & \\
              &\multicolumn{16}{l}{\underbrace{\hspace{4em}}_{I}\ \ \underbrace{\hspace{4em}}_{J}\ \ \underbrace{\hspace{4em}}_{I}\ \ \underbrace{\hspace{4em}}_{I}\hspace{3.5em} \underbrace{\hspace{4em}}_{I}} &&
        \end{array}.
    \end{equation*}
    Then the absolute match effect can be identified since the rank is the same as the number of absolute match effect parameters.

    Finally, we prove that absolute match effect and fixed effects are identifiable with restrictions $\sum_i \mu_{i j}=\sum_j \mu_{i j}=\sum_i \alpha_i=\sum_j \beta_j=0$.
    Each row in the coefficient matrix for restrictions $\sum_i \alpha_i=\sum_j \beta_j=0$, $R_2$ and $R_3$ defined as
    \begin{equation*}
        \begin{array}{cc}
             R_2=& \begin{pmatrix}
                 1 & \cdots & 1 & 0 & \cdots & 0 & 0 & \cdots & 0
             \end{pmatrix}\\
             R_3=& \begin{pmatrix}
                 0 & \cdots & 0 & 1 & \cdots & 1 & 0 & \cdots & 0
             \end{pmatrix} \\
             & \multicolumn{1}{l}{\hspace{1em}\underbrace{\hspace{3.5em}}_{I}\hspace{1.2em}\underbrace{\hspace{3.5em}}_{J}\hspace{1.2em}\underbrace{\hspace{3.5em}}_{IJ}} 
        \end{array}
    \end{equation*}
    are independent from every row in $A,B,C$ and $R_1$ and they have rank of $1$. 
    This is because the number of $a$'s in a row which is not zero is 2 while at least $I$ elements remains not zero by adding the rows in $R_2$, and similar for $R_3$ independence.
    Therefore, $rank\begin{pmatrix}
        [A^T & B^T & C^T & R_1^T & R_2^T & R_3^T]^T
    \end{pmatrix}=IJ+I+J$ which equals to the number of parameters, and thus absolute match effect and fixed effect parameters are identifiable.
\end{proof}

\subsection{Illustrative example}

Consider an illustrative example with $I=2$ and $J=3$. The relative match effect is estimated as
\begin{align*}
    \mu'_1 = \mu_{11}+\mu_{22}-\mu_{12}-\mu_{21} \\
    \mu'_2 = \mu_{11} +\mu_{23} -\mu_{13} -\mu_{21}
\end{align*}
and all the other relative match effect is calculated from $\mu'_1$ and $\mu'_2$. Without zero sum restrictions, absolute match effect $\mu_{ij}$ is not identifiable, but with restrictions $\sum_i\mu_{ij}=\sum_j\mu_{ij}=0$, the match effect is calculated as\footnote{You can derive $\mu_{11}$ from $\mu'_1+\mu'_2=2\mu_{11}+\mu_{22}+\mu_{23}-\mu_{12}-\mu_{13}-2\mu_{21}=6\mu_{11}$ by using the restrictions. And $\mu_{11}=-\mu_{21}=(\mu'_1+\mu'_2)/6$. Then, from the equations, you get $\mu_{12}=-\mu_{21}=(2\mu_{11}-\mu'_1)/2$ and $\mu_{13}=-\mu_{23}=(2\mu_{11}-\mu'_2)/2$. }
\begin{align*}
     \mu_{11} &= \frac{\mu'_{1}}{6} + \frac{\mu'_{2}}{6}\\
     \mu_{12} &= - \frac{\mu'_{1}}{3} + \frac{\mu'_{2}}{6}\\
     \mu_{13} &= \frac{\mu'_{1}}{6} - \frac{\mu'_{2}}{3}\\
     \mu_{21} &= - \frac{\mu'_{1}}{6} - \frac{\mu'_{2}}{6}\\
     \mu_{22} &= \frac{\mu'_{1}}{3} - \frac{\mu'_{2}}{6}\\
     \mu_{23} &= - \frac{\mu'_{1}}{6} + \frac{\mu'_{2}}{3}.
\end{align*}

\end{document}